\documentclass[12pt]{article}
\pdfoutput=1

\usepackage{amsmath,amssymb,amscd}
\usepackage{listings}
\usepackage{caption}
\usepackage{dsfont}
\usepackage{slashed}
\usepackage{color}

\usepackage[pdftex]{graphicx}
\usepackage{epstopdf}
\usepackage{subfigure}
\usepackage{epsfig}
\usepackage{listings}
\usepackage{caption}
\usepackage{cite}

\usepackage{multirow}

\newcommand{\bea}{\begin{align}}
\newcommand{\eea}{\end{align}}
\newcommand{\beq}{\begin{equation}}
\newcommand{\eeq}{\end{equation}}
\newcommand{\nbea}{\begin{align*}}
\newcommand{\neea}{\end{align*}}
\newcommand{\nbeq}{\begin{equation*}}
\newcommand{\neeq}{\end{equation*}}
\newcommand{\bear}{\begin{eqnarray}}  
\newcommand{\eear}{\end{eqnarray}}  

 \def\lra#1{\overset{\text{\scriptsize$\leftrightarrow$}}{#1}}

\numberwithin{equation}{section}

\begin{document}

\begin{titlepage}

\pagestyle{empty}

\baselineskip=21pt
{\small
\rightline{KCL-PH-TH/2015-47, LCTS/2015-35, CERN-PH-TH/2015-244}
\rightline{Cavendish-HEP-15/09, DAMTP-2015-62}
}
\vskip 0.8in

\begin{center}

{\Large {\bf Sensitivities of Prospective Future $e^+ e^-$ Colliders to Decoupled New Physics}}

\vskip 0.6in

 {\bf John~Ellis}$^{1,2}$
and {\bf Tevong~You}$^{1,3}$

\vskip 0.4in

{\small {\it
$^1${Theoretical Particle Physics and Cosmology Group, Physics Department, \\
King's College London, London WC2R 2LS, UK}\\
\vspace{0.25cm}
$^2${TH Division, Physics Department, CERN, CH-1211 Geneva 23, Switzerland} \\
\vspace{0.25cm}
$^3${Cavendish Laboratory, University of Cambridge, J.J. Thomson Avenue, \\
Cambridge, CB3 0HE, UK; \\
\vspace{-0.25cm}
DAMTP, University of Cambridge, Wilberforce Road, Cambridge, CB3 0WA, UK}
}}

\vskip 0.6in

{\bf Abstract}

\end{center}

\baselineskip=18pt \noindent


{

We explore the indirect sensitivities to decoupled new physics of prospective precision electroweak measurements, 
triple-gauge-coupling measurements and Higgs physics at future $e^+e^-$ colliders, with emphasis on the ILC250 and FCC-ee. 
The Standard Model effective field theory (SM EFT) is adopted as a model-independent approach for relating 
experimental precision projections to the scale of new physics, and we present prospective constraints on the 
Wilson coefficients of dimension-6 operators. We find that in a marginalised fit ILC250 EWPT measurements may be sensitive to 
new physics scales $\Lambda = \mathcal{O}(10)$~TeV, and FCC-ee
EWPT measurements may be sensitive to $\Lambda = \mathcal{O}(30)$~TeV. The prospective sensitivities
of Higgs and TGC measurements at the ILC250 (FCC-ee) are to $\Lambda = \mathcal{O}(1)$~TeV
($\Lambda = \mathcal{O}(2)$~TeV).
}


\vskip 0.8in

\leftline{October 2015}

\end{titlepage}

\newpage


\section{Introduction}

At the time of writing, all measurements of the known particles persist in remaining consistent with their
properties as predicted in the Standard Model (SM), and there are no convincing
signals at the LHC of any particles beyond those in the SM. In particular,
the couplings of the Higgs boson~\cite{Eureka} have recently been analysed
using the combined ATLAS and CMS data, with no signs of new physics~\cite{ATLASCMScombination}. Under these circumstances, it is natural
to assume that it is `the' SM Higgs boson, and suppose that any new physics must involve massive particles that are decoupled from physics
at the energies explored so far~\cite{decoupling}. A powerful tool for analysing such models of physics beyond the
SM is provided by the SM Effective Field Theory (SM EFT), which parametrises possible new
physics via a systematic expansion in a series of higher-dimensional operators composed of SM fields~\cite{buchmullerwyler, GIMR}.
The most important r\^ole in this approach is played by operators of dimension 6~\footnote{The unique dimension-5 operator is the well-known 
Weinberg neutrino-mass operator~\cite{weinbergoperator}. See Refs.~\cite{dim7} and \cite{dim8} for a classification of dimension-7 and -8 operators.}, 
whose matching to ultraviolet (UV) models is greatly simplified by the universal one-loop effective action when 
such operators are loop-induced~\cite{universal}.

The SM EFT has already been used in several analyses~\cite{earlyeft, HanSkiba, eftconstraints, ciuchinietal, PomarolRiva, ESY, falkowskiriva, berthiertrott, flavourfulEFT, EFThiggsreview, higgslegacyEFT, topquarkEFT, TGCEFT} of the available data from the LHC and previous
accelerators including LEP and the SLC, which set the standard for electroweak precision tests 
(EWPTs)~\footnote{Different operator bases in the literature may be translated between each other using the {\tt Rosetta} tool~\cite{rosetta}.}. 
As reviewed in Section~2 of this paper, there are certain (combinations of) dimension-6 operators whose coefficients are
particularly tightly constrained by these EWPTs. On the other hand, the coefficients of other (combinations of)
operators are constrained by other measurements, including Higgs physics and triple-gauge couplings (TGCs).
Together with Ver\'onica Sanz, we have previously published a global analysis of dimension-6 operators in the SM EFT~\cite{ESY},
providing 95\% CL ranges for their coefficients, both when each operator is switched on individually and when
marginalising over the possible coefficients of all contributing operators~\footnote{For other similar global fits to dimension-6 operators, 
see for example, Refs.~\cite{eftconstraints, PomarolRiva, higgslegacyEFT}.}. 

There is currently growing interest in the physics accessible to possible future $e^+ e^-$ colliders that
would continue the studies made with LEP and the SLC to higher energies and or 
luminosities~\cite{ILCwhitepaper, ILCquantum, ILC, ILCop, TLEPdesigndoc, futurestudies}. One of
the primary objectives of such machines will be make detailed studies of the Higgs boson and its interactions,
with other possible elements of their physics programmes including studies at the $Z$ peak with very high
luminosities, studies of $W^+ W^-$ production close to threshold and above, measurements near the
${\bar t} t$ threshold and, of course, searches for possible new particles. 

In this paper we explore the implications within the SM EFT of the high-precision physics possible 
with relatively low-energy $e^+ e^-$ colliders, considering in particular the ILC running at 250 GeV with an
integrated luminosity of 250~$\text{fb}^{-1}$~\cite{ILC}, the scenario we call ILC250, and FCC-ee with 10~$\text{ab}^{-1}$
at a centre-of mass energy of 240 GeV~\cite{TLEPdesigndoc}, both accompanied by lower-energy running at the $Z$ peak and
the $W^+ W^-$ threshold. We do not consider the possibilities for producing directly new particles, which are relatively
limited at these centre-of-mass energies.

In Section~2 of this paper we review briefly relevant aspects of the SM EFT, identifying the operators of
dimension 6 that are most relevant for the observables we consider. We then consider in Section~3 the formalism
we use for analysing prospective measurements of electroweak precision measurements, exhibiting the
corrections to SM predictions for EWPTs that one finds at first order in the SM EFT coefficients. This is
followed by analyses of the prospective constraints on these coefficients that could be provided by ILC250
and FCC-ee measurements. We then present a corresponding discussion of possible contributions within
the SM EFT to Higgs physics and TGC measurements, as well as analyses of the sensitivities to the corresponding
of ILC250 and FCC-ee. As we discuss, the prospective constraints on some (combinations of) operator
coefficients are so tight that they may as well be set to zero in the analyses of Higgs physics and TGCs.

When translated into the effective mass scales $\Lambda$ to which the prospective measurements are able to reach,
we find that ILC250 EWPT measurements could be sensitive to $\Lambda = \mathcal{O}(10)$~TeV, and FCC-ee
EWPT measurements could be sensitive to $\Lambda = \mathcal{O}(30)$~TeV, when marginalised over the effects of all relevant dimension-6 operators. The corresponding sensitivities
of Higgs and TGC measurements at the ILC250 (FCC-ee) are to $\Lambda = \mathcal{O}(1)$~TeV
($\Lambda = \mathcal{O}(2)$~TeV).

\section{The Standard Model Effective Field Theory}

In the Standard Model Effective Field Theory (SM EFT) the renormalisable interactions in the SM
are supplemented by higher-dimensional operators. These are composed of all possible combinations of 
SM fields that respect the $SU(3)_c \times SU(2)_L \times U(1)_Y$ gauge symmetries and Lorentz invariance, 
with the leading lepton-number-conserving effects parametrised by dimension $d \ge 6$ operators with unknown
Wilson coefficients that could be generated by decoupled new physics beyond the SM, assuming that
these also respect the SM gauge symmetries. According to
the decoupling assumption, the effects of operators with dimensions $d > 6$ are sub-leading, so we consider
just the dimension-6 SM EFT Lagrangian
\begin{equation}
\mathcal{L}_\text{SMEFT} = \mathcal{L}_\text{SM} + \sum_i \frac{c_i}{\Lambda^2}\mathcal{O}_i 	\, ,
\label{SMEFT}
\end{equation}
where the $\mathcal{O}_i$ are the dimension-6 operators in the basis of Ref.~\cite{PomarolRiva} that we adopt here,
$\Lambda$ represents the scale of new physics, and the coefficients $c_i$ depend on the details of its structure. 
The operators relevant for the observables included in our fits are listed in Table \ref{tab:dim6},
where we assume CP conservation and a flavour-blind structure for the operators involving SM fermions~\footnote{For studies that relax some of the flavour assumptions, see for example Refs.~\cite{flavourfulEFT, topquarkEFT}.}. 

\begin{table}[h]
{\small
\begin{center}
\begin{tabular}
{ | c | c | c | } 
\hline
EWPTs & Higgs Physics & TGCs \\
\hline
\multicolumn{3}{| c |}{ ${\mathcal O}_W=\frac{ig}{2}\left( H^\dagger  \sigma^a \lra {D^\mu} H \right )D^\nu  W_{\mu \nu}^a$  } \\
\hline
\multicolumn{2}{| c |}{ ${\mathcal O}_B=\frac{ig'}{2}\left( H^\dagger  \lra {D^\mu} H \right )\partial^\nu  B_{\mu \nu}$  } & ${\mathcal O}_{3W}= g \frac{\epsilon_{abc}}{3!} W^{a\, \nu}_{\mu}W^{b}_{\nu\rho}W^{c\, \rho\mu}$	\\
\hline
${\cal O}_T=\frac{1}{2}\left (H^\dagger {\lra{D}_\mu} H\right)^2$ & \multicolumn{2}{| c |}{ ${\mathcal O}_{HW}=i g(D^\mu H)^\dagger\sigma^a(D^\nu H)W^a_{\mu\nu}$ } \\
\hline
$\mathcal{O}_{LL}^{(3)\, l}=( \bar L_L \sigma^a\gamma^\mu L_L)\, (\bar L_L \sigma^a\gamma_\mu L_L)$ & \multicolumn{2}{| c |}{$ {\mathcal O}_{HB}=i g^\prime(D^\mu H)^\dagger(D^\nu H)B_{\mu\nu}$ } \\
\hline
${\mathcal O}_R^e = (i H^\dagger {\lra { D_\mu}} H)( \bar e_R\gamma^\mu e_R)$ & ${\mathcal O}_{g}=g_s^2 |H|^2 G_{\mu\nu}^A G^{A\mu\nu}$ &  \\
\hline
${\cal O}_{R}^u = (i H^\dagger {\lra { D_\mu}} H)( \bar u_R\gamma^\mu u_R)$ & ${\mathcal O}_{\gamma}={g}^{\prime 2} |H|^2 B_{\mu\nu}B^{\mu\nu}$ & \\
\hline
${\cal O}_{R}^d = (i H^\dagger {\lra { D_\mu}} H)( \bar d_R\gamma^\mu d_R)$ & ${\mathcal O}_H=\frac{1}{2}(\partial^\mu |H|^2)^2$ & \\
\hline
${\cal O}_{L}^{(3)\, q}=(i H^\dagger \sigma^a {\lra { D_\mu}} H)( \bar Q_L\sigma^a\gamma^\mu Q_L)$  & ${\mathcal O}_{f}   =y_f |H|^2    \bar{F}_L H^{(c)} f_R + \text{h.c.}$ & \\
\hline
${\cal O}_{L}^q=(i H^\dagger {\lra { D_\mu}} H)( \bar Q_L\gamma^\mu Q_L)$  & $\mathcal{O}_6 = \lambda|H|^6$ & \\ 
\hline
\end{tabular}
\end{center}
}
\caption{\it List of CP-even dimension-6 operators in our chosen basis~\cite{PomarolRiva}, noting in each case
the categories of observables that place the strongest constraints on the operator or its linear combinations with other operators.}
\label{tab:dim6}
\end{table}

The high-sensitivity electroweak precision tests (EWPTs), particularly those using the leptonic subset of $Z$-pole observables, 
impose the strongest constraints on the following dimension-6 operators:
\begin{equation}
\mathcal{L}_\text{dim-6}^\text{EWPT} \supset \frac{1}{2}\frac{(\bar{c}_W + \bar{c}_B)}{m_W^2} (\mathcal{O}_W + \mathcal{O}_B) + \frac{\bar{c}_T}{v^2}\mathcal{O}_T + \frac{\bar{c}^{(3)l}_{LL}}{v^2}\mathcal{O}^{(3)l}_{LL} + \frac{\bar{c}^e_R}{v^2}\mathcal{O}^e_R 	\, ,
\label{EWPTs}
\end{equation}
where we introduce coefficients $\bar{c}_i$ whose normalisations differ from those in (\ref{SMEFT})
by squared ratios of the electroweak scale to the nominal new-physics scale $\Lambda$:
\begin{equation}
\bar{c}_i = c_i \frac{M^2}{\Lambda^2} \, ,
\label{cbarc}
\end{equation}
where $M \equiv v, m_W$ depending on the operator. 

On the other hand, the dimension-6 operators and their linear combinations that affect Higgs physics and measurements of 
triple-gauge couplings (TGCs) in our fits are given by
\begin{align}
\mathcal{L}_\text{dim-6}^\text{Higgs+TGC} &\supset \frac{1}{2}\frac{(\bar{c}_W - \bar{c}_B)}{m_W^2}(\mathcal{O}_W - \mathcal{O}_B) +\frac{\bar{c}_{HW}}{m_W^2}\mathcal{O}_{HW} + \frac{\bar{c}_{HB}}{m_W^2}\mathcal{O}_{HB} + \frac{\bar{c}_g}{m_W^2}\mathcal{O}_{g} + \frac{\bar{c}_\gamma}{m_W^2}\mathcal{O}_\gamma  \nonumber \\
&\quad \quad + \frac{\bar{c}_H}{v^2}\mathcal{O}_H + \frac{\bar{c}_f}{v^2}\mathcal{O}_f 	\, .
\label{HTGCs}
\end{align}
Since the linear combination $\bar{c}_W + \bar{c}_B$ is potentially constrained very strongly by EWPTs,
we set $\bar{c}_B = -\bar{c}_W$ in the fits to Higgs physics and the TGCs. 

We note that the coefficients constrained in our fit correspond to those defined at the electroweak scale, $c_i \equiv c_i(v)$, which can be related to $c_i(\Lambda)$ at the matching scale by RGE running~\cite{RGE}. We neglect dimension-8 operators in our analysis,
as well as four-fermion operators (other than $\bar{c}^{(3)l}_{LL}$ that modifies the input parameter $G_F$),
whose effects on $Z$-pole measurements are formally of the same order as dimension-8 operators due to the 
lack of linear interference terms with the SM amplitudes~\cite{HanSkiba}. The effects of these operators
and other omitted theory uncertainties may be important for $\Lambda \lesssim 3$ TeV~\cite{berthiertrott} but, as we will see in the next Section,
the UV cut-off scale for future electroweak precision measurements can be assumed to be beyond this.

\section{Electroweak Precision Tests}

We use in our analyses of the electroweak precision tests (EWPTs) the $W$ mass and the following $Z$-peak pseudo-observables:
\begin{align*}
\Gamma_Z &= \Gamma_\text{had} + 3\Gamma_l + 3\Gamma_\nu \quad , \quad
R_l = \frac{\Gamma_\text{had}}{\Gamma_l}	\quad , \quad
R_q = \frac{\Gamma_q}{\Gamma_\text{had}}	\, , \\
\sigma_\text{had} &= 12\pi\frac{\Gamma_e\Gamma_\text{had}}{\hat{m}_Z^2\Gamma_Z^2}	\quad , \quad
A^f_{FB} = \frac{3}{4}A_e A_f 	\quad , \quad
m_W = c_W m_Z	\, .
\end{align*}
These are functions of the decay widths and asymmetries:
\begin{align*}
\Gamma_f &= \frac{\sqrt{2} G_F m_Z^2 \hat{m}_Z}{6\pi}\left(g_{f_L}^2 + g_{f_R}^2 \right)	\, ,\\
A_f &= \frac{g_{f_L}^2 - g_{f_R}^2}{g_{f_L}^2 + g_{f_R}^2}	\, ,
\end{align*}
which depend in turn upon modifications to the $Z\bar{f}f$ couplings:
\begin{equation*}
g_{f_L} = g^{SM}_{f_L} + \delta g_{f_L} \, , \quad
g_{f_R} = g^{SM}_{f_R} + \delta g_{f_R} \, ,
\end{equation*} 
where $g^{SM}_f = T^3_f - Q_f s^2_W$. 
These observables receive direct contributions from $\bar{c}^e_R, \bar{c}^u_R, \bar{c}^d_R, \bar{c}^q_L$ and $\bar{c}^{(3)q}_L$ through the 
following coupling modifications~\footnote{They
also depend on the coefficients $\bar{c}^L_L$ and $\bar{c}^{(3)L}_L$ of operators that are eliminated in the basis
we use~\cite{PomarolRiva}.}:
\begin{align}
\xi_{g^l_R} &\supset -\frac{1}{2}\frac{c^l_R}{g^l_R}	\quad , \quad \xi_{g^l_L} \supset 0 	\, , \\
\xi_{g^q_R} &\supset  -\frac{1}{2}\frac{c^q_R}{g^q_R}	\quad , \quad \xi_{g^q_L} \supset \frac{T^3_q c^{(3)q}_L - \frac{1}{2}c^q_L}{g^q_L}	\, ,
\label{shifts}
\end{align}
where we have defined the fractional shifts $\xi_X \equiv \delta X / X$, and we use the symbol $\supset$
to indicate that there are further shifts from other dimension-6 operators. The decay widths and asymmetries are then modified as follows:
\begin{align*}
\xi_{\Gamma_f} &\supset 2\frac{g_{f_L}^2 \xi_{g_{f_L}} + g_{f_R}^2 \xi_{g_{f_R}}}{g_{f_L}^2 + g_{f_R}^2}	\quad , \quad
\xi_{A_f} = 4\frac{g_{f_L}^2g_{f_R}^2}{g_{f_L}^4 - g_{f_R}^4} \left(\xi_{g_{f_L}} - \xi_{g_{f_R}} \right)	\, .
\end{align*}
There are also indirect corrections from the four-fermion operator $\bar{c}^{(3)l}_{LL}$, which modifies the input observable $G_F$ so that
\begin{align*}
\xi_{\Gamma_f} &\supset \xi_{G_F}	\quad , \quad 
\xi_{g^f_{L,R}} \supset \frac{-Q_f s^2_W}{T^3_f - Q_f s^2_W} \xi_{s^2_W}	\quad , \quad
\xi_{m_W} \supset -\frac{1}{2}\frac{s_W^2}{c_W^2}\xi_{s_W^2}		\, .
\end{align*}
Since $s^2_W = \frac{1}{2} - \frac{1}{2}\sqrt{1 - \frac{4\pi\alpha}{\sqrt{2}G_F m_Z^2}}$, there is a dependence of the
weak mixing angle $\theta_W$ on the modifications to $G_F$:
\begin{align*}
\xi_{s^2_W} \supset -\frac{c^2_W}{c_{2W}}\xi_{G_F}	\, .
\end{align*}
Finally, there are also indirect corrections from the oblique corrections
$\hat S \equiv \bar{c}_W + \bar{c}_B$ and $\hat T \equiv \bar{c}_T$ arising from contributions to self-energies $\delta \pi_{VV}$:
\begin{align*}
\delta \pi_{ZZ} &= -\hat{T} + 2\hat{S} s^2_W	\quad , \quad \delta \pi^\prime_{ZZ} = 2\hat{S} s^2_W	\, , \\
\delta \pi_{\gamma Z} &= -\hat{S} c_{2W}t_W	\quad , \quad \delta^\prime{\gamma\gamma} = -2\hat{S}s^2_W	\, ,
\end{align*}
where the self-energies $\pi_{VV}$ are defined as 
\begin{align*}
\pi_{ZZ} &\equiv \frac{\pi_{ZZ}(m_Z^2)}{m_Z^2}	\quad , \quad \pi^\prime_{ZZ} \equiv \lim_{q^2\to m_Z^2} \frac{\pi_{ZZ}(q^2) - \pi_{ZZ}(m_Z^2)}{q^2-m_Z^2}	\, , \\
\pi_{\gamma Z} &\equiv  \frac{\pi_{\gamma Z}(m_Z^2)}{m_Z^2}		\quad , \quad \pi^\prime_{\gamma\gamma} \equiv \lim_{q^2\to 0} \frac{\pi_{\gamma\gamma}(q^2) - \pi_{\gamma\gamma}(0)}{q^2}	\, , \\
\pi_{WW} &\equiv \frac{\pi_{WW}(m_W^2)}{m_W^2} 	\quad , \quad \pi^0_{WW} \equiv \frac{\pi_{WW}(0)}{m_W^2}	\, .
\end{align*}
These modifications affect the EWPTs through the corrections
\begin{align*}
m_Z^2 &= (m_Z^2)^0 (1 + \pi_{ZZ})	\quad , \quad G_F = G_F^0(1 - \pi_{WW}^0)	\quad , \quad \alpha(m_Z) = \alpha^0(m_Z)(1 + \pi^\prime_{\gamma\gamma})	\, , \\
m_W^2 &= (m_W^2)^0 (1 + \pi_{WW})	\quad , \quad \sin^2{\theta^f_\text{eff}} = s^2_W \left(1 - \frac{c_W}{s_W}\pi_{\gamma Z}\right)	\, .
\end{align*}
Using these results and the definition of $s^2_W$ gives
\begin{equation*}
\xi_{s^2_W} = \frac{c^2_W}{c_{2W}}\left( -\delta\pi^\prime_{\gamma\gamma} + \delta\pi_{ZZ} - \delta\pi^0_{WW} - \frac{c_{2W}}{s_W c_W}\delta\pi_{\gamma Z}\right)	\, .
\end{equation*}
Similarly, including all the previously-calculated corrections, we find for the decay width and $W$ mass  that 
\begin{align*}
\xi_{\Gamma_f} &=  \delta\pi^\prime_{ZZ} - \delta\pi_{ZZ} + \xi_{G_F}	\, , \\
\xi_{m_W} &= -\frac{1}{2}\frac{s_W^2}{c_W^2}\xi_{s_W^2} + \frac{1}{2}\delta\pi_{ZZ} + \frac{1}{2}\delta\pi_{WW}	\, .
\end{align*}

Focusing on the leptonic subset of observables, we may summarise numerically the dependences of the observables on 
the dimension-6 operator coefficients using the above tree-level expressions for the observables as follows:
\begin{align}
\xi_{\Gamma_Z} &= -2.69 \bar{c}^{(3)l}_{LL} - 0.19 \bar{c}^e_R + 1.35 \bar{c}_T - 0.90 \bar{c}^+_V	\, , \nonumber \\
\xi_{\sigma^0_\text{had}} &= 0.054 \bar{c}^{(3)l}_{LL} - 1.46 \bar{c}^e_R - 0.03 \bar{c}_T + 0.07 \bar{c}^+_V  \, , \nonumber \\
\xi_{R_e} &= -0.56 \bar{c}^{(3)l}_{LL} + 1.84 \bar{c}^e_R + 0.28 \bar{c}_T - 0.73 \bar{c}^+_V	\, , \nonumber \\
\xi_{A^e_{FB}} &= -71.38 \bar{c}^{(3)l}_{LL} + 28.89 \bar{c}^e_R + 35.69 \bar{c}_T - 92.90 \bar{c}^+_V \, , \nonumber \\
\xi_{m_W} &= -0.43\bar{c}^{(3)l}_{LL} + 0.72 \bar{c}_T - 1.02 \bar{c}^+_V \, ,  \nonumber \\
\xi_{A_e} &= -35.70 \bar{c}^{(3)l}_{LL} + 14.44 \bar{c}^e_R + 17.84 \bar{c}_T - 46.45 \bar{c}^+_V	\, .
\label{eq:EWPTdim6}
\end{align}
Here we have used ${s^2_W}|^\text{SM} \equiv 0.23162$, corresponding to the value obtained when relating the EWPT observables to the 
input parameters for the SM alone, to the highest theoretical precision available~\cite{wellszhang}. 

However, since we are neglecting the one-loop contributions 
from dimension-6 operators this precision is formally of a higher order in the calculations of the expansion coefficients,
and we may equally well use ${s^2_W}|^\text{tree} \equiv 0.21221$, which is the value obtained using the tree-level expressions 
to relate the EWPT observables to the input parameters~\cite{falkowskiriva}. Varying $s^2_W$ between these values can therefore
give an indication of the importance of these higher-order effects and is responsible, for example,
for the differences between the current EWPT limits given in~\cite{ESY} and in~\cite{PomarolRiva}.
Including consistently the effects of dimension-6 operators at the loop level would require going beyond the
tree level when calculating the expansion coefficients in (\ref{eq:EWPTdim6}). The importance of this omission
can also be estimated by calculating numerically the parametric dependences of observables using {\tt ZFITTER}~\cite{ZFITTER}, as in~\cite{wellszhang}, 
which includes the higher-loop contributions of input parameter modifications in the SM but still neglects the 
full loop contributions of the dimension-6 operators~\footnote{For some studies of including dimension-6 operators at the loop level see,
for example, Ref.~\cite{NLOEFT}.}. A complete study including the effects of dimension-6 operators at loop level is 
beyond the scope of this note. 

\begin{figure}[h!]
\begin{center} 
\includegraphics[scale=0.5]{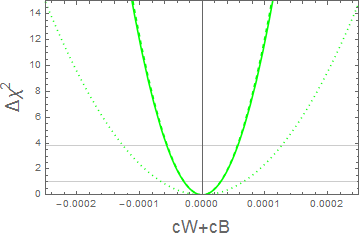}
\includegraphics[scale=0.5]{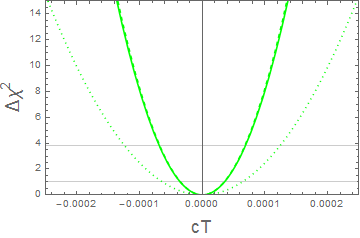}	\\
\includegraphics[scale=0.5]{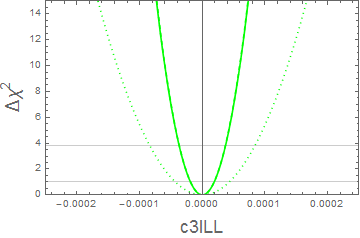}
\includegraphics[scale=0.5]{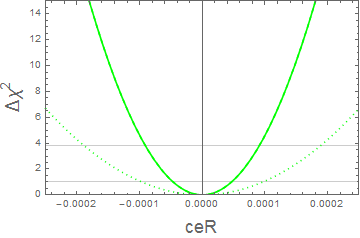}
\caption{\it Prospective constraints on individual operator coefficients from ILC EWPT measurements.
The projected $\Delta\chi^2$ for each operator affecting the leptonic subset of EWPT observables when 
switched on one at a time with the others set to zero, incorporating the prospective ILC measurements~\protect\cite{ILCquantum}.
The solid (dotted) lines are for dimension-6 contributions computed using the tree-level expressions with 
${s^2_W}|^\text{SM}$ $({s^2_W}|^\text{tree})$, whereas the dashed lines (indistinguishable here
from the solid lines) are computed numerically using {\tt ZFITTER} in the expansion framework of~\protect\cite{wellszhang}.   }
\label{fig:ILC_EWPT_individual}
\end{center}
\end{figure}

\subsection{EWPT constraints from the ILC}

\begin{figure}[h!]
\begin{center} 
\includegraphics[scale=0.5]{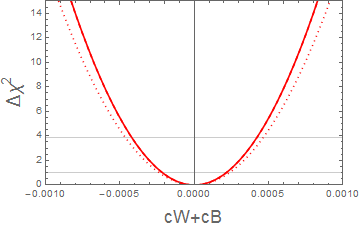}
\includegraphics[scale=0.5]{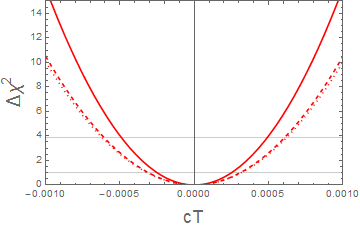}	\\
\includegraphics[scale=0.5]{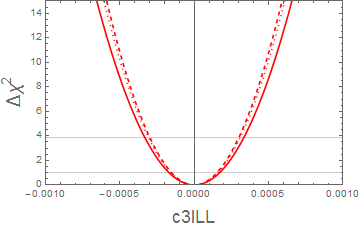}
\includegraphics[scale=0.5]{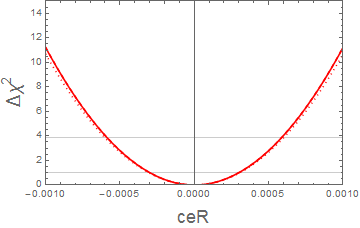}
\caption{\it Prospective marginalised constraints on operator coefficients from ILC EWPT measurements.
The projected $\Delta\chi^2$ for each operator affecting the leptonic subset of EWPT observables when 
when they are all allowed to vary simultaneously, incorporating the prospective ILC measurements~\protect\cite{ILCquantum}.
The solid (dotted) lines are for dimension-6 contributions computed using the tree-level expressions with 
${s^2_W}|^\text{SM}$ $({s^2_W}|^\text{tree})$, whereas the dashed lines
are computed numerically using {\tt ZFITTER} in the expansion framework of~\protect\cite{wellszhang}. }
\label{fig:ILC_EWPT_marg}
\end{center}
\end{figure}

\begin{figure}[h!]
\begin{center} 
\includegraphics[scale=0.5]{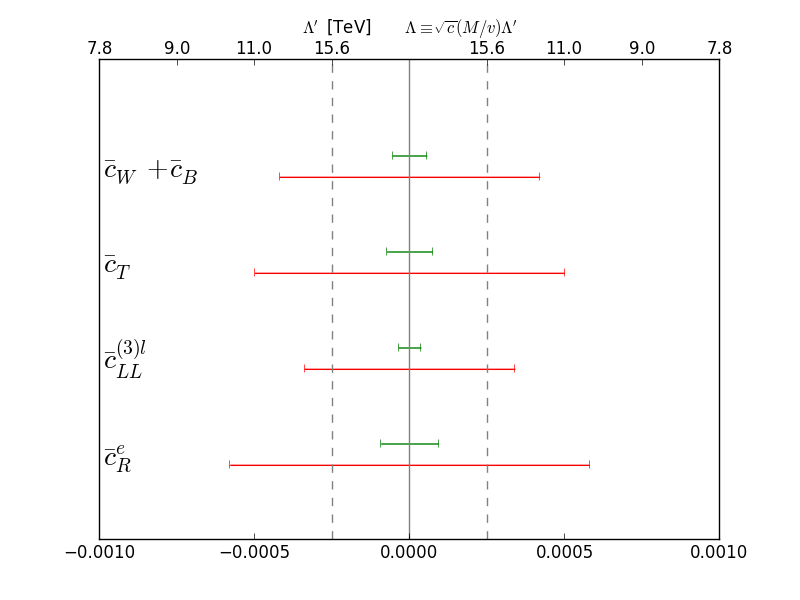}	
\caption{\it Summary plot of the individual (green) and marginalised (red) 95 \% CL limits on dimension-6 operators at ILC. The upper axis denotes the cut-off scale $\Lambda$ when $c\sim 1$. For $\bar{c}_W + \bar{c}_B$ with an operator normalisation of $M^2 = m_W^2$ instead of $v^2$ this should be read as divided by $\sim 3$. }
\label{fig:ILC_EWPT_summary}
\end{center}
\end{figure}

We take the following 1-$\sigma$ ILC experimental errors for the observables $\{ m_W, \Gamma_Z, R_l, A_e \}$ from~\cite{ILCquantum}:
\begin{equation}
\sigma_{m_W} = 0.005 \text{ GeV}	\quad , \quad
\sigma_{\Gamma_Z} = 0.001 \text{ GeV}	\quad , \quad
\sigma_{R_l} = 0.01	\quad , \quad
\sigma_{A_e} = 0.0001	\, .
\label{eq:ILCsigmas}
\end{equation}
We neglect theoretical uncertainties in the SM predictions for these quantities for the purposes of our analysis. For reference, we recall that
the current theoretical uncertainty in the SM prediction of $m_W$ is estimated to be 4 MeV, which is
potentially reducible to $\sim 1$ MeV when higher-order contributions are calculated in the future~\cite{ILCquantum}. We assume that
this will occur within the time-scale of the measurements considered here.

The result of a $\chi^2$ fit to the prospective ILC EWPT measurements assuming Gaussian errors,
switching on each operator individually and
setting the others to zero, is shown in Fig.~\ref{fig:ILC_EWPT_individual}. The solid and dotted lines denote the dimension-6 
contributions to the EWPT observables using the tree-level expressions with ${s^2_W}|^\text{SM}$ and ${s^2_W}|^\text{tree}$ respectively. 
The dashed line represents the result of using {\tt ZFITTER} in the expansion formalism of~\cite{wellszhang} to calculate the 
dimension-6 contributions including the higher-order corrections to the parametric dependences. 
In this case we see results that are practically identical to the solid lines, with 95\% CL limits at the $\sim 10^{-5}$ level.

The marginalised $\chi^2$ for a fit allowing all four dimension-6 operators to vary simultaneously is displayed in Fig.~\ref{fig:ILC_EWPT_marg}, 
where we see that the limits extend to $\sim 10^{-4}$. In this case some small differences can be observed between the 
solid and dashed lines for $\bar{c}^e_R$ and $\bar{c}^{(3)l}_{LL}$, whereas none are visible for $\bar{c}^e_R$ and $\bar{c}_W + \bar{c}_B$. 
In the case of $\bar{c}^e_R$, this is due to the fact that both calculations use ${s^2_W}|^\text{SM}$, and at tree level $\bar{c}^e_R$ does not 
modify the input parameters whose higher-order contributions are taken into account in the dashed lines, so that no difference is expected.
 
The projected individual and marginalised 95\% CL ILC uncertainties in the four dimension-6 operators are shown in green and red respectively in 
Fig.~\ref{fig:ILC_EWPT_summary}. The upper axis converts the limits on the barred coefficients $\bar{c}_i$ to an energy scale in TeV when 
$c_i \sim 1$ (corresponding to an $\mathcal{O}(1)$ new physics coupling) and the operator normalisation is $v^2/\Lambda^2$. 
For $\bar{c}_W + \bar{c}_B$, whose operator normalisation is $m_W^2 / \Lambda^2$, the energy scale is effectively divided by $\sim 3$. 

\subsection{EWPT constraints from FCC-ee}

\begin{figure}[h!]
\begin{center} 
\includegraphics[scale=0.7]{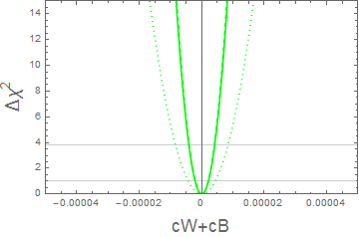}
\includegraphics[scale=0.7]{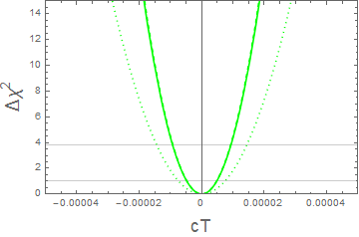}	\\
\includegraphics[scale=0.7]{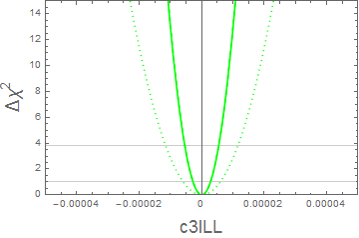}
\includegraphics[scale=0.7]{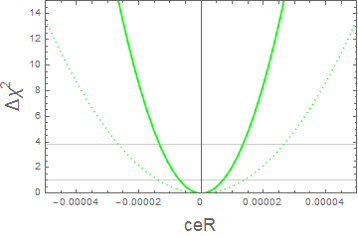}
\caption{\it  Prospective constraints on individual operator coefficients from FCC-ee EWPT measurements.
The projected $\Delta\chi^2$ for each operator affecting the leptonic subset of EWPT observables when 
switched on one at a time with the others set to zero, incorporating the prospective FCC-ee measurements~\protect\cite{TLEPblondel,TLEPdesigndoc}.
The solid (dotted) lines are for dimension-6 contributions computed using the tree-level expressions with 
${s^2_W}|^\text{SM}$ $({s^2_W}|^\text{tree})$, whereas the dashed lines (indistinguishable here
from the solid lines) are computed numerically using {\tt ZFITTER} in the expansion framework of~\protect\cite{wellszhang}. }
\label{fig:TLEP_EWPT_individual}
\end{center}
\end{figure}

\begin{figure}[h!]
\begin{center} 
\includegraphics[scale=0.524]{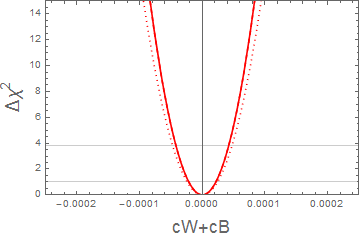}
\includegraphics[scale=0.7]{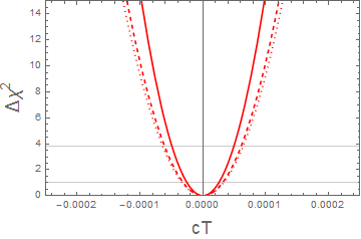}	\\
\includegraphics[scale=0.7]{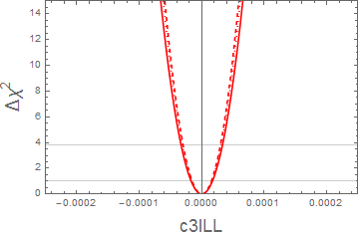}
\includegraphics[scale=0.7]{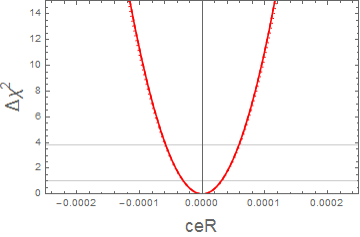}
\caption{\it Prospective marginalised constraints on operator coefficients from FCC-ee EWPT measurements.
The projected $\Delta\chi^2$ is shown for each operator affecting the leptonic subset of EWPT observables when 
when they are all allowed to vary simultaneously, incorporating the prospective FCC-ee measurements~~\protect\cite{TLEPblondel,TLEPdesigndoc}.
The solid (dotted) lines are for dimension-6 contributions computed using the tree-level expressions with 
${s^2_W}|^\text{SM}$ $({s^2_W}|^\text{tree})$, whereas the dashed lines
are computed numerically using {\tt ZFITTER} in the expansion framework of~\protect\cite{wellszhang}.}
\label{fig:TLEP_EWPT_marg}
\end{center}
\end{figure}

\begin{figure}[h!]
\begin{center} 
\includegraphics[scale=0.6]{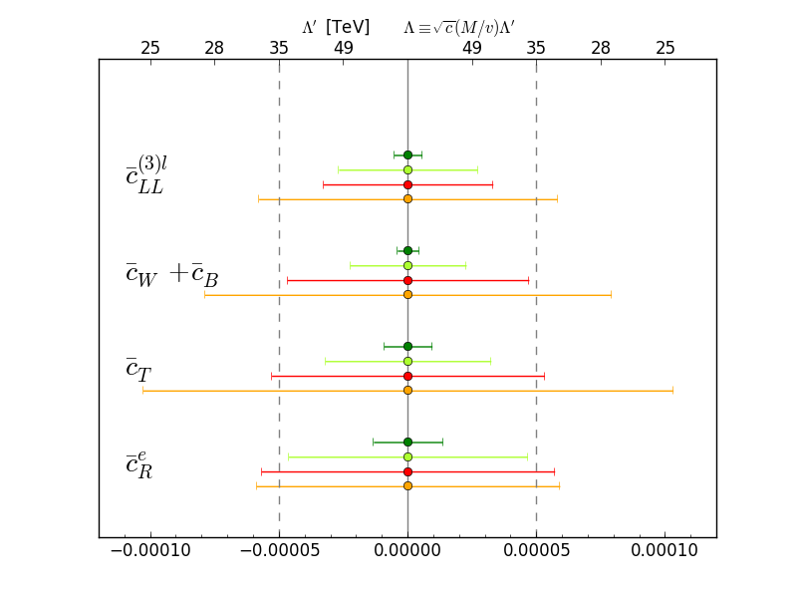}	
\caption{\it Projected 95 \% CL limits at $\text{FCC-ee}^{10 \text{ab}^{-1}}_\text{240 GeV}$ for the leptonic subset of operators affecting EWPTs. The individual (marginalised) bounds are coloured in dark green (red). The effects of theoretical uncertainties are included in light green (orange) for the individual (marginalised) fits.  }
\label{fig:TLEP_EWPT_summary}
\end{center}
\end{figure}

We now analyse the prospective sensitivity of the FCC-ee, using the same set of four observables as for the ILC,
but with the experimental errors given in~\cite{TLEPblondel}, that are based on~\cite{TLEPdesigndoc}, 
\begin{equation}
\sigma_{m_W} = 0.0005	\text{ GeV} \quad , \quad
\sigma_{\Gamma_Z} = 0.0001 \text{ GeV}	\quad , \quad
\sigma_{R_l} = 0.001	\quad , \quad
\sigma_{A_e} = 0.000015	\, .
\label{FCCeeuncertainties}
\end{equation}
These errors are dominated by the systematic uncertainties, and we neglect again the theoretical uncertainties, 
so as to indicate the potential sensitivity of the experimental reach alone. 

Fig.~\ref{fig:TLEP_EWPT_individual} shows the $\chi^2$ contributions from individual fits switching on the operators one at a time,
and Fig.~\ref{fig:TLEP_EWPT_marg} varies them simultaneously before marginalising over the other operators.
As previously, the solid and dashed lines denote the $\chi^2$ contributions of dimension-6 operators calculated with $s^2_W|^\text{SM}$
using the tree-level expressions and the {\tt ZFITTER} expansion coefficients of~\cite{wellszhang}, respectively. 
They are indistinguishable in the individual fits. whereas the marginalised fit shows some small variations. The dotted lines
calculated using the tree-level expressions with $s^2_W|^\text{tree}$ exhibit larger variations in the individual fit.

The prospective FCC-ee
95\% CL constraints are summarised in dark green (red) for the individual (marginalised) limits in Fig.~\ref{fig:TLEP_EWPT_summary}. 
Even in the marginalised case, the barred coefficients are constrained at the $\mathcal{O}(10^{-5})$ level,
which translates to an indirect sensitivity in the tens of TeV, modulo the effects of weak or strong coupling in the new physics being integrated at tree or loop level. 

With statistical and systematic uncertainties reduced to the levels shown in (\ref{FCCeeuncertainties}),
the limiting factors in interpreting the data may well be the theoretical uncertainties that we have neglected.
Attaining the optimal sensitivity to indirect effects from physics beyond the SM
will require reducing these theoretical uncertainties, the effects of which can be estimated by adding in quadrature the 
projections from~\cite{mishimaslides}, 
\begin{equation*}
\sigma^\text{th}_{\Gamma_Z} = 0.0001 \text{ GeV} \quad , \quad \sigma^\text{th}_{m_W} = 0.001 \text{ GeV} \quad , \quad \sigma^\text{th}_{A_e} = 0.000118	\, .
\end{equation*}
The current theoretical uncertainties are 4 MeV for the $W$ mass, 0.5 MeV for the $Z$ decay width and $37 \times 10^{-5}$ for $A_l$,
which could be reduced to the above estimates by future three-loop level calculations~\cite{mishimaslides}. The resulting individual and marginalised 
95 \% CL constraints are shown in Fig.~\ref{fig:TLEP_EWPT_summary} in light green and orange respectively.

\section{Higgs and Triple-Gauge Couplings}

The Higgs production mechanism we consider for future $e^+ e^-$ colliders is associated $Z + H$ production. 
The dependence on the dimension-6 coefficients of the cross section for this process at a centre-of-mass energy $\sqrt{s} \sim 250$ GeV
can be expressed via a rescaling factor 
relative to the SM prediction that was calculated in~\cite{craigetal}, which can be translated into our basis and normalisation as
\begin{align*}
\frac{\delta \sigma_\text{VH}}{\sigma_\text{VH}} &\approx 1 + 1.98 \bar{c}^{(3)l}_{LL} + 1.16 \bar{c}_B + 1.55 \bar{c}_\gamma - 12.6 \bar{c}^e_R - 
 0.99 \bar{c}_H \\
 &- 0.77 \bar{c}_{HB} + 7.74 \bar{c}_{HW} - 0.661 \bar{c}_T + 19.3 \bar{c}_W	\, ,
\end{align*}
where we set $\bar{c}^e_R = \bar{c}^{(3)l}_{LL} = 0$ and $\bar{c}_B = -\bar{c}_W$,
due to the strong EWPT constraints on these combinations of operators described previously. 
The numerical dependences of the Higgs branching ratios on the dimension-6 operator coefficients are provided in~\cite{eHDECAY}.

For triple-gauge couplings (TGCs) we use the $e^+ e^- \to W^+ W^-$ cross-section rescalings at $\sqrt{s} = 200$ and 500 GeV calculated 
in~\cite{wellszhangTGC}~\footnote{For other studies of dimension-6 operators in TGCs see for example Refs.~\cite{TGCEFT, CEPCTGC, CPviolatingTGC, higgsTGC, bobethhaisch}}. Using the integration-by-parts identity
\begin{equation*}
\mathcal{O}_B = \mathcal{O}_{HB} + \frac{1}{4}\mathcal{O}_{BB} + \frac{1}{4}\mathcal{O}_{WB} 	\, ,
\end{equation*}
the expressions in~\cite{wellszhangTGC} are translated into our basis and operator normalisation as
\begin{align*}
\left. \frac{\delta \sigma_{WW}}{\sigma_{WW}}\right|_{500 \text{GeV}} &\approx 0.47(\bar{c}_{HW} + \bar{c}_W) + 0.52(\bar{c}_{HW} + \bar{c}_{HB}) + 0.18\bar{c}_{3W} - 0.76(\bar{c}_W + \bar{c}_B) + 22.30\bar{c}_T  \, , \\
\left. \frac{\delta \sigma_{WW}}{\sigma_{WW}}\right|_{200 \text{GeV}} &\approx 0.05(\bar{c}_{HW} + \bar{c}_W) + 0.095(\bar{c}_{HW} + \bar{c}_{HB}) + 0.05\bar{c}_{3W} - 0.74(\bar{c}_W + \bar{c}_B) + 14.93\bar{c}_T 	\, .
\end{align*}
We see that the cross-section dependence on the effects of dimension-6 operators rises with the energy of diboson production,
which makes this an important channel for constraining the SM EFT. As in the expressions for the Higgscouplings, 
we set here $\bar{c}_W + \bar{c}_B$ and $\bar{c}_T$ to zero as these are more strongly constrained by EWPTs. 
Since the ILC projections are given only for each TGC anomalous couplings individually, and none are available for FCC-ee, 
we use for both ILC and FCC-ee an $\mathcal{O}(10^{-4})$ experimental sensitivity, corresponding to an
improvement by two orders of magnitude over the per-cent measurements at LEP2 as estimated in Ref.~\cite{ILCop}. 

\subsection{Higgs and TGC constraints from the ILC}

The scenario we consider is the ILC running at 250 GeV with the standard luminosity of 250 $\text{fb}^{-1}$,
which we call ILC250. The error projections for the different Higgs channels are taken from Table 5.4 of~\cite{ILCwhitepaper},
and the TGCs are included as described above. The prospective $Z\gamma$ Higgs branching ratio
measurement is not reported, so we conservatively take the error on this to be 100\%. 

\begin{figure}[h!]
\begin{center} 
\includegraphics[scale=0.5]{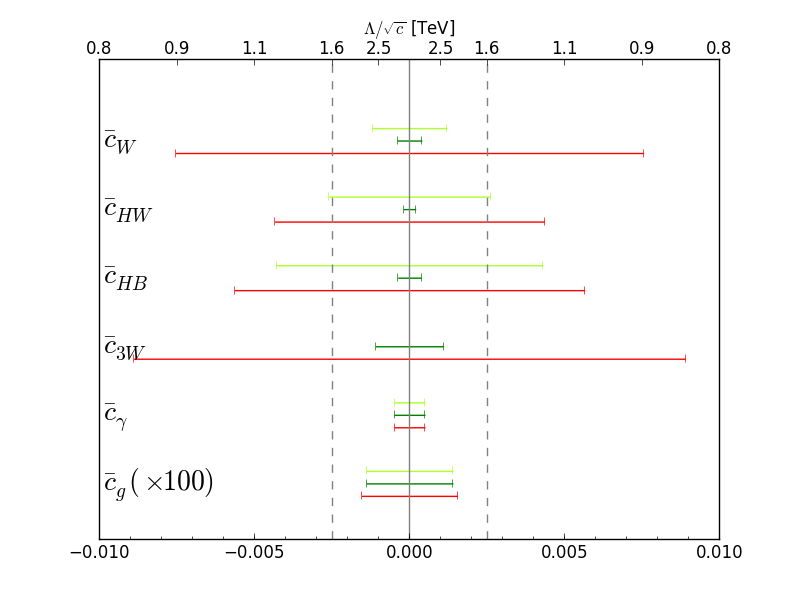}	
\caption{\it  Summary of the 95 \% CL constraints on dimension-6 operator coefficients affecting Higgs and TGC observables at ILC250.
The green bars indicate individual fits switching on one coefficient at a time, with light green using Higgs measurements only and 
dark green also including TGCs. The marginalised constraints are denoted by red bars. The upper x-axis should be rescaled by $\times 3$ $(\times 10)$ for 
$\bar{c}_\gamma$ $(\bar{c}_g)$. }
\label{fig:ILC_HiggsTGC_summary}
\end{center}
\end{figure}

We perform a 9-parameter $\chi^2$ fit to the operator coefficients $\{ \bar{c}_W, \bar{c}_{HW}, \bar{c}_{HB}, \bar{c}_{3W}, \bar{c}_\gamma, 
\bar{c}_g,$ $\bar{c}_H, \bar{c}_u$ and $\bar{c}_d \}$ using a Metropolis-Hastings algorithm. The resulting widths of the 95\% CL constraints
centred on zero are summarised in Fig.~\ref{fig:ILC_HiggsTGC_summary}, where we omit for clarity the coefficients 
$\bar{c}_H, \bar{c}_u, \bar{c}_d$, whose limits are an order of magnitude worse. The dark green (red) bounds denote the individual (marginalised) fits. 

We see that even in the marginalised case the limits are at the $\sim 10^{-3}$ level,
which indicates a sensitivity that begins to probe the TeV scale. This is to be contrasted with the limits on these coefficients from the LHC,
which are currently at the per-cent level. The importance of including TGCs can also be seen by their effect on the individual limits when 
removing them from the fit, as shown in light green in Fig.~\ref{fig:ILC_HiggsTGC_summary}. 
The $\bar{c}_g$ $(\bar{c}_\gamma)$ coefficient is multiplied by $100$ $(10)$ to be visible on the same scale, 
which translates into multiplying the upper x-axis by $\sim 10$ $(3)$. The actual scales that may be indirectly probed are of course 
dependent on the new physics couplings and potential loop suppression factors. 

\subsection{Higgs and TGC constraints from FCC-ee}

We consider now the FCC-ee running at 240 GeV with the standard scenario of $10$ $\text{ab}^{-1}$ of luminosity. 
The estimated errors for each Higgs channel are summarised in Table 4 of~\cite{TLEPdesigndoc}, 
and we assume the same TGC and $Z\gamma$ Higgs branching ratio projections as for the ILC. 

\begin{figure}[h!]
\begin{center} 
\includegraphics[scale=0.5]{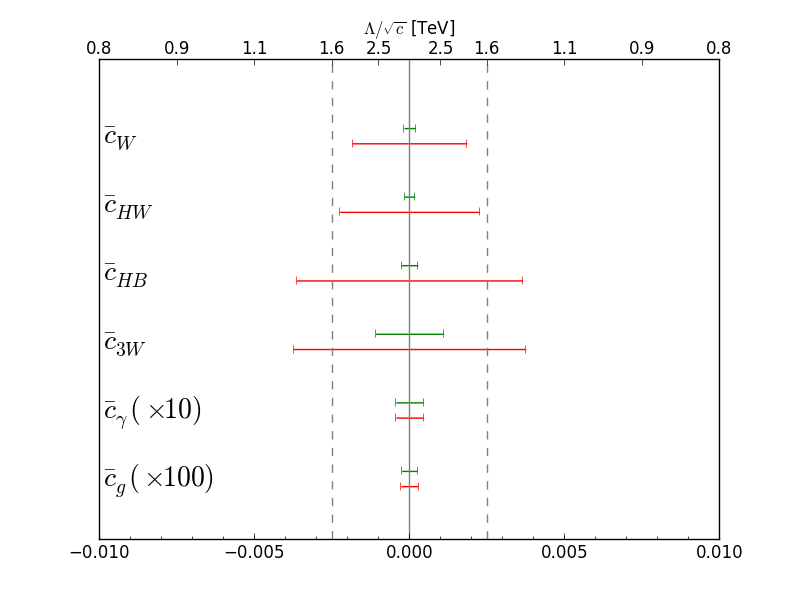}	
\caption{\it  Summary of the 95 \% CL limits on dimension-6 operator coefficients affecting Higgs and TGC observables at FCC-ee. 
The individual (marginalised) limits are shown in green (red). The upper x-axis should be rescaled by $\times 3$ $(\times 10)$ for 
$\bar{c}_\gamma$ $(\bar{c}_g)$. }
\label{fig:TLEP_HiggsTGC_summary}
\end{center}
\end{figure}

The results of the $\chi^2$ fit are shown in Fig.~\ref{fig:TLEP_HiggsTGC_summary},
with the same colour codings as for the ILC250 case. We see that the 95 \% CL limits are now well into the TeV range. 
The $\bar{c}_g$ $(\bar{c}_\gamma)$ coefficient is again multiplied by $100$ $(10)$ to be visible on the same scale, 
and we recall that the actual scales that may be indirectly probed
depend on the new physics couplings and potential loop suppression factors. 

\section{Conclusions}

\begin{figure}[h!]
\begin{center} 
\includegraphics[scale=0.3885]{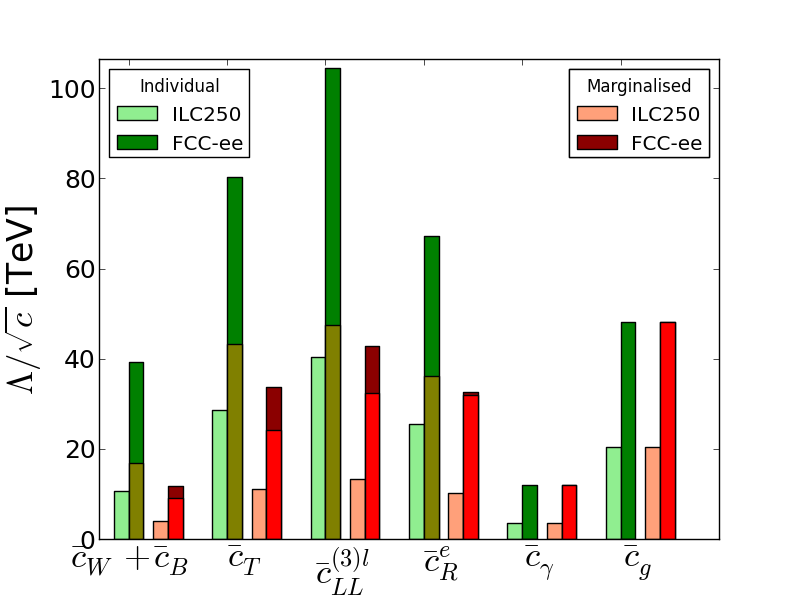}
\includegraphics[scale=0.3885]{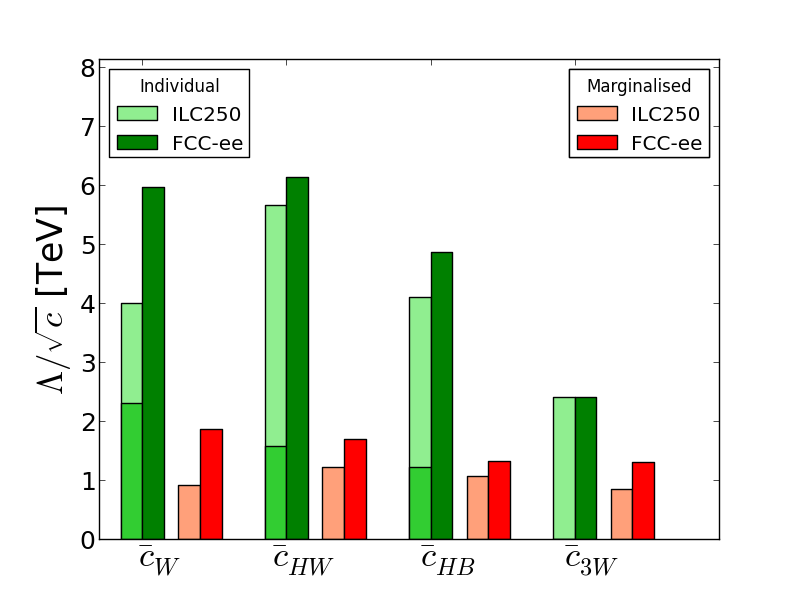}	
\caption{\it Summary of the reaches for the dimension-6 operator coefficients with TeV scale sensitivity, when switched on individually (green) and when marginalised (red), from projected precision measurements at the ILC250 (lighter shades) and FCC-ee (darker shades). The left plot shows the operators that are most strongly constrained by EWPTs and Higgs physics, where the different shades of dark green and dark red represent the effects of EWPT theoretical uncertainties at FCC-ee. The right plot is constrained by Higgs physics and TGCs, and the different shades of light green demonstrate the improved sensitivity when TGCs are added at ILC250. }
\label{fig:ILC_vs_FCCee}
\end{center}
\end{figure}

Until fundamentally new particles beyond the SM are discovered, the SM EFT may be viewed as the Fermi theory of the $21^\text{st}$ century.
It is the effective low-energy theory given all experimentally established degrees of freedom, 
and the objective is to measure a non-zero Wilson coefficient that might indicate the structure of new physics. 
The systematic classification of possible effects from decoupled new physics makes this an attractive framework for 
characterising the impacts of measurements across the SM as a whole~\footnote{It is worth mentioning that the 
possible breakdown of the SM EFT assumptions in specific measurements is not a weakness, but a strength of the approach, 
as it could provide a consistency check that informs the way forward in investigating any new physics effects.}. 

The importance of improving precision tests of the SM, in particular in the Higgs sector, strongly motivates the construction of a future lepton collider. 
Such proposals include the ILC and FCC-ee, as well as the Chinese collider CEPC. 
One may then ask how the improved precision of measurements at these machines translates into the scale of heavy new physics
to which we shall be indirectly sensitive. The SM EFT provides a relatively model-independent way to address this question.

We have shown in this paper that the prospective sensitivities of possible future $e^+ e^-$ colliders
extend to $\Lambda = \mathcal{O}(30)$~TeV in the case of EWPTs at FCC-ee, $\Lambda = \mathcal{O}(10)$~TeV
in the case of EWPTs at ILC250, $\Lambda = \mathcal{O}(2)$~TeV in the case of Higgs and TGC measurements
at FCC-ee, and $\Lambda = \mathcal{O}(1)$~TeV in the case of Higgs and TGC measurements at ILC250. 
These estimates are for the more conservative marginalised limits. The individual fits, 
assuming only one operator affects a given set of observables at a time, provides an upper bound on the potential reach. 
These results are summarised in Fig.~\ref{fig:ILC_vs_FCCee}. We expect that higher-energy runs of the ILC 
would improve the sensitivity to new physics via Higgs and TGC
measurements, but improving its sensitivity to new physics via EWPTs would require higher luminosity at the $Z$
peak and near the $W^+ W^-$ threshold. In this respect, the capabilities of the CEPC or the ILC with upgraded luminosity would lie between those of the ILC250 and FCC-ee.

As noted earlier in this paper, there are significant uncertainties in our analysis of EWPTs associated with the absence of a complete
loop treatment of the SM EFT contributions to them. However, these uncertainties are unlikely to 
affect qualitatively the results of our analyses of Higgs physics and TGCs. Also as noted earlier, full exploitation of the potential of ILC and particularly
FCC-ee measurements will require a new generation of precision electroweak and QCD loop calculations to match the
statistical and other experimental uncertainties. These calculations, together with the inclusion of SM EFT theoretical errors, will certainly require a concerted and substantial theoretical effort.


\section*{Acknowledgements}
TY thanks Adam Falkowski and Francesco Riva for helpful consistency checks, Michael Trott for related discussions, and Markus Klute for the invitation to participate in the FCC-ee Higgs physics workshop at CERN where this work was completed.  
The work of JE was supported partly by the London Centre
for Terauniverse Studies (LCTS), using funding from the European Research Council via the Advanced Investigator
Grant 26732, and partly by the STFC Grants ST/J002798/1 and ST/L000326/1.
The work of TY was supported initially by a
Graduate Teaching Assistantship from King's College London
and now by a Junior Research Fellowship from Gonville and Caius College, Cambridge.

 \providecommand{\href}[2]{#2}\begingroup\raggedright


\begin{thebibliography}{10}

  \bibitem{Eureka}
G.~Aad {\it et al.}  [ATLAS Collaboration],
  Phys.\ Lett.\ B {\bf 716} (2012) 1
  [arXiv:1207.7214 [hep-ex]];
S.~Chatrchyan {\it et al.}  [CMS Collaboration],
  Phys.\ Lett.\ B {\bf 716} (2012) 30
  [arXiv:1207.7235 [hep-ex]].

\bibitem{ATLASCMScombination}
ATLAS Collaboration, ATLAS-CONF-2015-044 (2015)

  \bibitem{decoupling}
  T. Appelquist and J. Carazzone, Phys. Rev. D 11 (1975) 285
  
  \bibitem{buchmullerwyler}
  W.~Buchmuller and D.~Wyler,
  Nucl.\ Phys.\ B {\bf 268} (1986) 621.
  
    \bibitem{GIMR}
    B.~Grzadkowski, M.~Iskrzynski, M.~Misiak and J.~Rosiek,
  JHEP {\bf 1010}, 085 (2010)
  [arXiv:1008.4884 [hep-ph]].
  
  
  \bibitem{weinbergoperator}
   S. Weinberg, 
   Phys.\ Rev.\ Lett.\  {\bf 43} (1979) 1566.
  
    \bibitem{dim7}
    L.~Lehman,
  Phys.\ Rev.\ D {\bf 90} (2014) 12,  125023
  [arXiv:1410.4193 [hep-ph]].
    
  \bibitem{dim8}
  L.~Lehman and A.~Martin,
  arXiv:1510.00372 [hep-ph].
 
 \bibitem{universal}
 B.~Henning, X.~Lu and H.~Murayama,
  arXiv:1412.1837 [hep-ph].
  A.~Drozd, J.~Ellis, J.~Qu\'evillon and T.~You,
  JHEP {\bf 1506} (2015) 028
  [arXiv:1504.02409 [hep-ph]] and in preparation.
  
    \bibitem{earlyeft}
    B. Grinstein and M. B. Wise, 
    Phys.Lett. B {\bf 265} (1991) 326–334.
  K.~Hagiwara, S.~Ishihara, R.~Szalapski and D.~Zeppenfeld,
  Phys.\ Rev.\ D {\bf 48}, 2182 (1993).
   K.~Hagiwara, R.~Szalapski and D.~Zeppenfeld,
  Phys.\ Lett.\ B {\bf 318}, 155 (1993)
  [hep-ph/9308347].
  
  \bibitem{HanSkiba}
Z.~Han and W.~Skiba,
  Phys.\ Rev.\ D {\bf 71} (2005) 075009
  [hep-ph/0412166].
  
  \bibitem{eftconstraints}
    T.~Corbett, O.~J.~P.~Ebol, J.~Gonzalez-Fraile and M.~C.~Gonzalez-Garcia,
arXiv:1211.4580 [hep-ph];
   B.~Dumont, S.~Fichet and G.~von Gersdorff,
  JHEP {\bf 1307}, 065 (2013)
  [arXiv:1304.3369 [hep-ph]].
  
    \bibitem{ciuchinietal}
  M.~Ciuchini, E.~Franco, S.~Mishima and L.~Silvestrini,
  JHEP {\bf 1308} (2013) 106
  [arXiv:1306.4644 [hep-ph]].
  M.~Ciuchini, E.~Franco, S.~Mishima, M.~Pierini, L.~Reina and L.~Silvestrini,
  arXiv:1410.6940 [hep-ph].
  
  \bibitem{PomarolRiva}
A.~Pomarol and F.~Riva,
  JHEP {\bf 1401} (2014) 151
  [arXiv:1308.2803 [hep-ph]].
  
  \bibitem{ESY}
J.~Ellis, V.~Sanz and T.~You,
  JHEP {\bf 1407} (2014) 036
  [arXiv:1404.3667 [hep-ph]].
  J.~Ellis, V.~Sanz and T.~You,
  JHEP {\bf 1503} (2015) 157
  [arXiv:1410.7703 [hep-ph]].
  
\bibitem{falkowskiriva}
   A.~Falkowski and F.~Riva,
  JHEP {\bf 1502} (2015) 039
  [arXiv:1411.0669 [hep-ph]].
  
    \bibitem{berthiertrott}
  L.~Berthier and M.~Trott,
  JHEP {\bf 1505} (2015) 024
  [arXiv:1502.02570 [hep-ph]].
   L.~Berthier and M.~Trott,
  arXiv:1508.05060 [hep-ph].
  
  \bibitem{flavourfulEFT}
  A.~Efrati, A.~Falkowski and Y.~Soreq,
  JHEP {\bf 1507} (2015) 018
  [arXiv:1503.07872 [hep-ph]].
  
    \bibitem{EFThiggsreview}
  A.~Falkowski,
  arXiv:1505.00046 [hep-ph].
  
  
  \bibitem{higgslegacyEFT}
  T.~Corbett, O.~J.~P.~Eboli, D.~Goncalves, J.~Gonzalez-Fraile, T.~Plehn and M.~Rauch,
  JHEP {\bf 1508} (2015) 156
  [arXiv:1505.05516 [hep-ph]].
  
  \bibitem{topquarkEFT}
  A.~Buckley, C.~Englert, J.~Ferrando, D.~J.~Miller, L.~Moore, M.~Russell and C.~D.~White,
  arXiv:1506.08845 [hep-ph].
  
  
  \bibitem{TGCEFT}
  A.~Falkowski, M.~Gonzalez-Alonso, A.~Greljo and D.~Marzocca,
  arXiv:1508.00581 [hep-ph].
  
   \bibitem{rosetta}
 A.~Falkowski, B.~Fuks, K.~Mawatari, K.~Mimasu, F.~Riva and V.~Sanz,
  arXiv:1508.05895 [hep-ph].
  
\bibitem{ILCwhitepaper}
    D.~M.~Asner {\it et al.},
  arXiv:1310.0763 [hep-ph].
  
  \bibitem{ILCquantum}
    A.~Freitas, K.~Hagiwara, S.~Heinemeyer, P.~Langacker, K.~Moenig, M.~Tanabashi and G.~W.~Wilson,
  arXiv:1307.3962 [hep-ph].
  
\bibitem{ILC}
  T.~Han, Z.~Liu, Z.~Qian and J.~Sayre,
  Phys.\ Rev.\ D {\bf 91} (2015) 113007
  [arXiv:1504.01399 [hep-ph]].
  G.~Moortgat-Pick {\it et al.},
  Eur.\ Phys.\ J.\ C {\bf 75} (2015) 8,  371
  [arXiv:1504.01726 [hep-ph]].
   K.~Fujii {\it et al.},
  arXiv:1506.05992 [hep-ex].
  
  \bibitem{ILCop}
  T.~Barklow, J.~Brau, K.~Fujii, J.~Gao, J.~List, N.~Walker and K.~Yokoya,
  arXiv:1506.07830 [hep-ex].
  
  
\bibitem{TLEPdesigndoc}
M.~Bicer {\it et al.} [TLEP Design Study Working Group Collaboration],
  JHEP {\bf 1401} (2014) 164
  [arXiv:1308.6176 [hep-ex]].
  
  
\bibitem{futurestudies}
M.~Baak {\it et al.},
  arXiv:1310.6708 [hep-ph].
   J.~Fan, M.~Reece and L.~T.~Wang,
  arXiv:1411.1054 [hep-ph].
  J.~Fan, M.~Reece and L.~T.~Wang,
  JHEP {\bf 1508} (2015) 152
  [arXiv:1412.3107 [hep-ph]].
   A.~Thamm, R.~Torre and A.~Wulzer,
  JHEP {\bf 1507} (2015) 100
  [arXiv:1502.01701 [hep-ph]].
  
  
  
\bibitem{RGE}
C.~Grojean, E.~E.~Jenkins, A.~V.~Manohar and M.~Trott,
  JHEP {\bf 1304} (2013) 016
  [arXiv:1301.2588 [hep-ph]].
  J.~Elias-Miró, J.~R.~Espinosa, E.~Masso and A.~Pomarol,
  JHEP {\bf 1308} (2013) 033
  [arXiv:1302.5661 [hep-ph]].
  J.~Elias-Miro, J.~R.~Espinosa, E.~Masso and A.~Pomarol,
  JHEP {\bf 1311} (2013) 066
  [arXiv:1308.1879 [hep-ph]].
  E.~E.~Jenkins, A.~V.~Manohar and M.~Trott,
  JHEP {\bf 1310} (2013) 087
  [arXiv:1308.2627 [hep-ph]].
  E.~E.~Jenkins, A.~V.~Manohar and M.~Trott,
  JHEP {\bf 1401} (2014) 035
  [arXiv:1310.4838 [hep-ph]].
  R.~Alonso, E.~E.~Jenkins, A.~V.~Manohar and M.~Trott,
  arXiv:1312.2014 [hep-ph].
    J.~Elias-Miro, C.~Grojean, R.~S.~Gupta and D.~Marzocca,
  arXiv:1312.2928 [hep-ph].
  R.~Alonso, H.~M.~Chang, E.~E.~Jenkins, A.~V.~Manohar and B.~Shotwell,
  Phys.\ Lett.\ B {\bf 734} (2014) 302
  [arXiv:1405.0486 [hep-ph]].
  
  
    \bibitem{wellszhang}
J.~D.~Wells and Z.~Zhang,
  Phys.\ Rev.\ D {\bf 90} (2014) 033006
  [arXiv:1406.6070 [hep-ph]].
  
  
    \bibitem{ZFITTER}
D.~Y.~Bardin, P.~Christova, M.~Jack, L.~Kalinovskaya, A.~Olchevski, S.~Riemann and T.~Riemann,
  Comput.\ Phys.\ Commun.\  {\bf 133} (2001) 229
  [hep-ph/9908433].
  
  
  \bibitem{NLOEFT}
  M.~Ghezzi, R.~Gomez-Ambrosio, G.~Passarino and S.~Uccirati,
  JHEP {\bf 1507} (2015) 175
  [arXiv:1505.03706 [hep-ph]].
   C.~Hartmann and M.~Trott,
  arXiv:1507.03568 [hep-ph].
    A.~David and G.~Passarino,
  arXiv:1510.00414 [hep-ph].
  
  \bibitem{TLEPblondel}
  A. Blondel, Exploring the Physics Frontier with Circular Colliders, Aspen, Colorado (USA), Jan. 31, 2015:
 {\tt http://http://indico.cern.ch/event/336571/}.
  
    \bibitem{mishimaslides}
   S. Mishima, 
   6th TLEP workshop, CERN, Oct. 16, 2013: {\tt http://indico.cern.ch/event/257713/session/1/contribution/30}.
  
    \bibitem{craigetal}
   N.~Craig, M.~Farina, M.~McCullough and M.~Perelstein,
  JHEP {\bf 1503}, 146 (2015)
  [arXiv:1411.0676 [hep-ph]].
  
  \bibitem{eHDECAY}
   R.~Contino, M.~Ghezzi, C.~Grojean, M.~Muhlleitner and M.~Spira,
  arXiv:1403.3381 [hep-ph].
  
  \bibitem{wellszhangTGC}
  J.~D.~Wells and Z.~Zhang,
  arXiv:1507.01594 [hep-ph].
  
  
  \bibitem{CEPCTGC}
L.~Bian, J.~Shu and Y.~Zhang,
  JHEP {\bf 1509} (2015) 206
  [arXiv:1507.02238 [hep-ph]].
  
  \bibitem{CPviolatingTGC}
  B.~Gripaios and D.~Sutherland,
  Phys.\ Rev.\ D {\bf 89} (2014) 7,  076004
  [arXiv:1309.7822 [hep-ph]].
  
\bibitem{higgsTGC}
T.~Corbett, O.~J.~P.~Eboli, J.~Gonzalez-Fraile and M.~C.~Gonzalez-Garcia,
  Phys.\ Rev.\ Lett.\  {\bf 111} (2013) 011801
  [arXiv:1304.1151 [hep-ph]].
  
    \bibitem{bobethhaisch}
  C.~Bobeth and U.~Haisch,
  JHEP {\bf 1509} (2015) 018
  [arXiv:1503.04829 [hep-ph]].
  


 
\end{thebibliography}
\end{document}